\documentclass[12pt]{article}
\usepackage[margin=1in]{geometry}
\usepackage{amsmath,amsthm,amssymb,amsfonts,setspace,enumerate,graphicx,tikz,hyperref,paralist,bbm,natbib,tikz, pgfplots}
\usepackage[margin=1cm,font=small,labelfont=bf]{caption}
\pgfplotsset{compat=1.17}
\usepackage[appendix=append]{apxproof} 

\renewcommand{\appendixsectionformat}[2]{Proofs}

\newcounter{thm}
\newcounter{lm}

\newtheorem{theorem}[thm]{Theorem}
\newtheorem{lemma}[lm]{Lemma}
\newtheorem{definition}{Definition}
\newtheorem{assumption}{Assumption}
\newtheorem{proposition}{Proposition}

\newtheorem{example}{Example}
\newtheorem*{example*}{Example}

\newcommand{\supp}{\operatorname{supp}}

\title{Flexible Learning via Noise Reduction}

\author{Peter Achim\footnote{Dept. of Economics, University of York. Email: peter.achim@york.ac.uk}\ \  \ \  Kemal Ozbek\footnote{Dept. of Economics, University of Southampton. Email: k.ozbek@soton.ac.uk}
}
\date{\today
\bigskip\bigskip}

\begin{document}

\maketitle

{\vspace{-1cm}\begin{center} --- PRELIMINARY AND INCOMPLETE --- \end{center}}

\bigskip
\begin{abstract}
    We develop a novel framework for costly information acquisition in which a decision-maker learns about an unobserved state by choosing a signal distribution, with the cost of information determined by the distribution of noise in the signal. We show that a natural set of axioms admits a unique integral representation of the cost function, and we establish the uniform dominance principle:  there always exists an optimal experiment that generates signals with uniform noise. The uniform dominance principle allows us to reduce the infinite-dimensional optimization problem of finding an optimal information structure to finding a single parameter that measures the level of noise. We show that an optimal experiment exists under natural conditions, and we characterize it using generalized first-order conditions that accommodate non-smooth payoff functions and decision rules. Finally, we demonstrate the tractability of our framework in a bilateral trade setting in which a buyer learns about product quality.
\end{abstract}

\section{Introduction}
\nosectionappendix

A growing body of work in economic theory investigates how decision-makers acquire information in uncertain environments when learning is costly. In many settings, the decision-maker faces a twofold challenge: choosing an optimal action and determining how much costly information to acquire about an uncertain state. A widely used framework for studying these issues is the model of rational inattention, introduced by \cite{sims2003implications}. In that framework, agents select the amount of information they wish to acquire by balancing the benefits of reduced uncertainty against the costs of processing information. Traditionally, these costs are quantified using the mutual information function—a measure of the reduction in uncertainty defined in terms of Shannon entropy \citep[e.g.][]{matvejka2015rational}.

Despite its success in individual decision-making problems, the mutual information-based framework faces significant challenges in interactive or strategic environments. In many such settings—such as bilateral trade—agents’ beliefs and actions are interdependent. As noted by \cite{denti2022experimental}, the conventional approach can lead to the “free at full information” (FFI) property, in which information costs vanish as uncertainty is nearly eliminated. This peculiar feature may result in counterintuitive outcomes, for example, scenarios in which buyers learn about sellers’ actions at no cost even in the absence of direct observation, thereby undermining the practical relevance of the model in strategic contexts.

To overcome these limitations, recent research has explored alternative approaches. \cite{denti2022experimental} advocates for an approach that shifts attention away from the posterior beliefs toward the statistical process that generates these beliefs, while \cite{de2017rationally} identified a set of intuitive properties that information cost functions should in general satisfy. Building on these insights, we develop a new framework in which experiments themselves are taken as primitives. In our approach, the decision-maker chooses an experiment by selecting a noise distribution that affects the signal about the true state.

In our setup, each experiment is characterized by a mapping from the true state to a signal generated by adding noise drawn from a chosen distribution. This design enables the decision-maker to directly control the level of noise and, hence, the precision of the signal. Greater attention (or effort) reduces noise and yields a more informative signal, but it also increases the cost of information acquisition. Conversely, allowing for a broader noise distribution reduces cost at the expense of informativeness. The cost of an experiment is defined as the expected cost of the noise—averaged over the prior distribution of the state—and it satisfies a set of natural axioms such as consistency, prior-independence (in the case of state-invariant experiments), linearity, continuity, and Blackwell monotonicity. These axioms ensure that more informative experiments are costlier and imply that the cost function admits a unique integral representation in terms of a measurable, continuous, and strictly unimodal noise cost function.

A key contribution of our work is the derivation of the \emph{uniform dominance principle}. We show that for any given experiment there exists another experiment—one that uses a uniform noise distribution across states—that generates a net benefit at least as high as the original experiment. This result is particularly important because it reduces the original infinite-dimensional problem of choosing among arbitrary noise distributions to a one-dimensional problem of selecting an optimal noise level. In other words, rather than searching over the full space of possible noise distributions, the decision-maker can focus on the simpler task of choosing the optimal spread of a uniform noise distribution. To accommodate potential non-smoothness in payoffs and decision rules, we derive generalized first-order conditions (in the sense of Clarke) that characterize the optimal experiment. These conditions capture the precise balance at which the marginal benefit of increased precision exactly offsets the marginal cost of reducing noise.

To illustrate the practical implications of our framework, we apply our analysis to a bilateral trade model in which a buyer learns about the quality of a product. In this application, our model yields a unique equilibrium outcome that contrasts sharply with the indeterminacy found in previous models based on mutual information costs (e.g., \cite{ravid2020ultimatum}). Moreover, by directly incorporating the statistical process of noise generation into the cost structure, our approach can be extended to a variety of other settings, including principal-agent problems and strategic decision-making scenarios.

\subsection*{Related Literature}

The foundational work of Sims (2003) introduced the rational inattention (RI) framework, which models decision-makers as optimally acquiring information while constrained by a processing cost quantified via mutual information (e.g., \cite{matvejka2015rational}). This approach has been widely adopted due to its tractability and empirical applicability (see, e.g, \cite{cdl22} for generalizations of this approach with posterior separable costs of information.) However, as emphasized by \cite{denti2022experimental}, the entropy-based cost function exhibits counterintuitive properties in equilibrium settings, particularly in games where beliefs are endogenous. Specifically, RI often results in the "free at full information" (FFI) property, which implies that as uncertainty is reduced, information acquisition costs vanish. This can lead to unrealistic equilibrium predictions, such as costless monitoring in strategic interactions.

To address these concerns, alternative formulations have been explored. \cite{ravid2020ultimatum} examines ultimatum bargaining under rational inattention and finds that RI generates excessive equilibrium multiplicity due to the degenerate cost of full information. He proposes a trembling-hand-like refinement, akin to Selten’s (1975) concept, which forces buyers to consider even low-probability deviations by sellers. This refinement eliminates inefficient equilibria in which buyers engage in costless off-path learning and restores uniqueness in equilibrium outcomes. When attention costs are sufficiently low, trade occurs, albeit with the buyer systematically overpaying for low-quality goods and underpaying for high-quality goods \cite{ravid2020ultimatum}. His results underscore the importance of refining equilibrium concepts to address the credibility of off-path beliefs.

An alternative to posterior-separable models is to treat experiments as the fundamental objects of choice, with costs assigned directly to them. This approach originates in Wald's (1945) statistical decision theory. \cite{morris2019wald} extends this perspective by linking sequential sampling models to ex-ante information costs. They demonstrate that all posterior-separable cost functions have an equivalent representation as solutions to a sequential learning problem, thereby bridging the gap between ex-ante and dynamic information acquisition models. Their results highlight the limitations of RI in dynamic settings and reinforce the necessity of modeling information acquisition explicitly through experiments.

\cite{pomatto2023cost} formalizes this approach further by providing an axiomatic foundation for experiment-based cost functions. They show that information costs can be represented as a function of the Kullback-Leibler divergence between experimental distributions. Unlike RI, this framework naturally avoids the FFI property and ensures that informative experiments always incur a positive cost, even in the limit of full information.

\cite{bloedel2020cost} characterize posterior-separable cost functions, which generalize mutual information by allowing costs to depend on posterior beliefs. They formalize the conditions under which posterior-separability holds and provide a sequential learning foundation for cost functions that depend on belief distributions rather than the underlying experiment. Their results illustrate that entropy-based cost functions are a specific instance of a broader class of models and highlight cases where uniform posterior separability (UPS) provides a more flexible yet consistent cost structure.

The implications of costly learning extend beyond individual decision-making into market and strategic contexts. \cite{mensch2022monopoly} analyze how costly learning shapes product quality choices in monopolistic settings. They show that when consumers endogenously acquire information about product quality, optimal product design must account for the structure of learning costs. Their model adopts a posterior-separable cost framework but highlights how learning constraints introduce new inefficiencies relative to standard pricing models.


\section{Model}

Consider a decision maker who must choose an action $a$ from a set of available actions $\mathcal A$ which we assume is a compact subset of some suitable
space with an underlying topology. Suppose that the decision maker
has some utility function $u(\theta,a)$ which depends continuously on her chosen action $a\in \mathcal A$ and on an uncertain state of nature $\theta \in \mathbb{R}$. The random variable $\theta$ is drawn from a distribution $F \in \Delta$, where $\Delta$ denotes the set of all probability distributions over the Borel space $(\mathbb{R}, \mathcal{B})$ represented by their cumulative distribution functions (c.d.f.). We endow the Borel space with the weak* topology, which says that a sequence of distributions ${G_n}$ converges to $G$ if and only if $\int f  dG_n \to \int f  dG$ for all bounded continuous functions $f : \mathbb{R} \to \mathbb{R}$. 


\subsection{Experiments}
The decision maker does not observe the state $\theta$ directly, but she can acquire information about it by performing a statistical experiment. An experiment generates a random signal $s$, which depends on the underlying state of nature $\theta \in \Theta\subseteq\mathbb{R}$ and a random noise variable $x \in \mathbb{R}$. In particular, each experiment $P\equiv\{ P_\theta \}$ assigns to each $\theta \in \mathbb{R}$ a probability distribution $P_\theta\in \Delta$ over the noise space. The relationship between the state, the noise, and the resulting signal is captured by a measurable signal function $\sigma: \mathbb{R} \times \mathbb{R} \to \mathbb{R}$, which maps each  realized pair $(\theta, x)$ to an observable signal $s\equiv \sigma(\theta, x) \in \mathbb{R}$. We let $S\equiv\sigma(\mathbb{R},\mathbb{R})$ denote the set of observable signals. We assume that $\sigma(\theta,x)$ is continuous in its arguments, that is, the functions $\sigma_{\theta}\equiv\sigma(\theta,.)$ and $\sigma_{x}\equiv\sigma(.,x)$ are continuous bijections on $\mathbb R$.

We call an experiment $P$ \textit{admissible} if it satisfies the following regularity conditions.
\begin{assumption}\label{ass:regular_P}
For each $\theta\in\mathbb R$, (i) $P_\theta$ is absolutely continuous with respect to the Lebesgue measure on $\mathbb R$, (ii) $\frac{dP_\theta}{dx}$ is an even function which is strictly unimodal at $0$.\footnote{A function $h$ is called even if $h(x)=h(-x)$ for all $x\in \mathbb R$. A function is strictly unimodal at zero if it is strictly increasing on $\mathbb R_-$ and strictly decreasing on $\mathbb R_+$.}

\end{assumption}
 Absolute continuity of $P_\theta$ ensures the existence of a density $\frac{dP_\theta}{dx}$ almost everywhere. Evenness and unimodality imply that the exiperiment produces an error that is evenly spread around the true state, and that the signal is an unbiased estimator. 

We call an experiment state-invariant if it assigns the same distribution over the noise regardless
of the state; that is if  $P_{\theta}=P_{\theta'}$ for all $\theta,\theta'\in\mathbb R$. Let $\bar{\mathcal E}$ denote the
set of state-invariant experiments. With some abuse of notation, for any $P\in \bar {\mathcal E}$, we use $P$ to denote the cdf for each state $\theta$, that is, $P_\theta=P$ for all $\theta\in \mathbb R$. For any given
$F\in\Delta $ and $P\in{\mathcal E}$, let $P^{F}\in{\bar{\mathcal E}}$ denote the state-invariant averaged experiment with $P^{F}(x)=\int_\mathbb R P_{\theta}(x)dF(\theta)$
for all $x\in\mathbb{R}$. Let $\mathcal{E}_u$ denote the set of uniform experiments and $\bar{\mathcal{E}}_u$ denote the set of state-invariant uniform experiments.

For any $P,P'\in{\mathcal E}$ and
$\alpha\in[0,1]$, we let $P\alpha P'$ denote the mixed experiment
where for each state, experiment $P_\theta$ is used with probability $\alpha$, and $P_\theta'$ with the complementary probability $1-\alpha$, i.e.,  for each $\theta$,
\[
(P\alpha P')_\theta=\alpha P_\theta+(1-\alpha)P'_\theta
\]

To compare experiments in terms of their informativeness in our framework, we introduce a modified version of the Blackwell order. The standard definition of the Blackwell order does not apply here, because we define experiments as  distributions over noise  rather than signals. The difference is that in contrast to signal, which derive their meaning from the statistical dependence on the underlying state, noise carries specific, deterministic information. To illustrate this, consider a garbling function that collapses all random draws to zero. If this garbling is applied to a signal distribution, the garbled signal becomes independent of the state, and thus is completely uninformative. In contrast, if this garbling is applied to a noise distribution, the garbled signal is in fact completely noise-free, and thus perfectly reveals the underlying state. To address this issue, we define a \textit{restricted} Blackwell order that only permits garbling functions to redistribute probability mass away from the center (zero) toward the tails. Formally, consider a stochastic kernel $K$, which is a map from $\mathbb R\times{\mathcal B}$
into $[0,1]$ such that for every $x\in \mathbb R$ and $E\in \mathcal{B}$, the set
function $K(x,\cdot)$ is a probability measure and the real-valued map
$K(.,E)$ is measurable. We say $K$ is restricted if $K(y,[-\delta,\delta])=0$ for
all $|y|>\delta$ for all $\delta\geq0$. In the following, we write $[x]$ in short to denote the event $(-\infty, x]$ for any $x\in \mathbb{R}$. For a given experiment $P=\{P_\theta\}$, we denote by $KP_\theta$ the cumulative distribution function given by
\[
KP_\theta(x)=\int_{y\in \mathbb{R}} K(y,[x])dP_\theta(y).
\]
We further write $KP=\{KP_\theta\}_{\theta \in \mathbb {R}}$.

\begin{definition}
    An experiment $P$ is  {(weakly) more informative} than $Q$  in the restricted Blackwell order (denoted $P \trianglerighteq Q$) if there exists a restricted kernel $K$ such that $Q=
    KP$. We say $P$ is strictly more informative than $Q$ in the restricted Blackwell order if $P \trianglerighteq Q$ but $Q \ntrianglerighteq P$.
\end{definition}



\subsection{Costs of information}
We assume that experiments are costly to process. Suppose that there
is a non-zero bounded function $C_F:\mathcal E\to\mathbb{R}_{+}$
such that for any given prior belief $F\in\Delta$,
the cost of processing an experiment $P\in{\mathcal E}$ is given
by $C_F(P)\geq0$. We assume the information cost function $C$
satisfies the following conditions:
\begin{assumption}\label{axm:cost_P}
Let $F, G \in \Delta$. 
\begin{enumerate}[$(i)$]
    \item Consistency: $C_F(P) = C_F( P^{F})$ for $P \in \mathcal{E}$.
    \item Prior-independence: $C_F(P) = C_G(P)$ for $P \in \bar{\mathcal E}$.
    \item \textit{Linearity}: $C_F( P) = C_F(P')$ implies $C_F(P \frac{1}{2} P'') = C_F(P' \frac{1}{2} P'')$ for $P, P', P'' \in \bar {\mathcal E}$.
    \item \textit{Continuity}: $P^{n} \to P$ implies $C_F( P^{n}) \to C_F(P)$ for $\{P^{n}\} \subset \bar {\mathcal E}$ and $P \in \bar {\mathcal E}$.
    \item \textit{Blackwell-Monotonicity}: $P \trianglerighteq P'$ implies $C_F(P) \geq C_F(P')$ and $P \triangleright  P'$ implies $C_F(P) > C_F(P')$ for $P, P' \in \bar {\mathcal{E}}$.
\end{enumerate}
\end{assumption}

Consistency says that the total cost of processing an experiment is an expectation over the costs for individual states. It ensures that if we aggregate the states into a state-invariant experiment (e.g., one that averages over states using the prior), the overall cost remains unchanged.  Prior independence says that if the experiment is state-invariant (i.e., the noise distribution is the same across all states), the cost of processing such an experiment should not depend on the prior, as the prior does not influence how the experiment operates. Linearity says that if two state-invariant experiments have equal costs, then a mixture of these with a third experiment in equal proportion will also yield equal costs.  Note that we could alternatively require that mixtures of any proportion must generate the same cost, but the condition we provide is weaker than that. Continuity ensures that the cost function $C$ behaves continuously, so that small changes in the experiment $P^n$ (approaching $P$) result in only small changes in cost. Finally, Blackwell-monotonicity captures the usual notion that acquiring more information generates a higher cost.

We now show that the conditions given in above axiom implies that
for any given belief $F\in\Delta$, the cost function has the following integral form. 
\begin{proposition}\label{prop:cost characterization}
    The cost function $C_{F}(.)$  satisfies  conditions $(i)$-$(iv)$ in Assumption \ref{axm:cost_P} for all $F\in \Delta$ if and only if there exists a measurable, continuous function $c:\mathbb R\to \mathbb R$ which is strictly unimodal at zero, such that 
\begin{equation}
C_{F}(P)=\int_{\theta\in\mathbb{R}} \left(\int_{x\in\mathbb{R}}c(x)dP_\theta(x) \right)dF(\theta)\label{eq:cost_P}
\end{equation}
for any $P \in \mathcal{E}$.
\end{proposition}

\begin{toappendix}
The following lemma is needed in the proof of Proposition \ref{prop:cost characterization}. 
\begin{lemma}\label{lm:garbled Q exists} Let $P\in \bar{\mathcal E}$. For any points $x_0$, $\hat x$, $x_1\in \mathbb R$ with $x_0<\hat x<x_1$ and $P(x_1)>P(x_0)$,  there exists $Q\in \bar {\mathcal E}$ such that $P\triangleright Q$ and
\begin{enumerate}
    \item $P(x)=Q(x)$ for all $x\in[0,x_0]\cup[x_1,\infty)$,
    \item $P(x)>Q(x)$ for all $x\in(x_0,\hat x)$,
    \item $P(x)<Q(x)$ for all $x\in[\hat x,x_1)$.
\end{enumerate}
    
\end{lemma}
\begin{proof}
Fix $\alpha\in (0,1)$. Consider a kernel $K$ which redistributes a share $\alpha$ of the probability mass from all points in $(x_0,\hat x)$ uniformly on $[\hat x,x_1)$ :
\begin{enumerate}[$(i)$]
    \item For $x \in [0, x_0] \cup (\hat x, \infty)$, let $K(y, E) = \delta_y(E)$ for all $E\in {\mathcal B}$, where $\delta_x$ is the Dirac  measure at $y$.
    \item For $x \in (x_0, \hat{x})$, set $    K(x, E) = (1 - \alpha) \delta_x(E) + \alpha \mu(E)$ for all $E\in {\mathcal B}$, where $\mu(E)$ denotes the uniform distribution:
        \[
        \mu(E) = \frac{\lambda(E \cap (\hat{x}, x_1))}{x_1 - \hat{x}}.
        \]
    Here, $\lambda(\cdot)$ denotes the Lebesgue measure.
\end{enumerate}
 It easy to verify that $K$ is a restricted Kernel, and the garbled experiment $Q=KP$ is given by
\[
Q(x)=\begin{cases} 
    P(x) &\text{ for $x\in [0,x_0)\cup [x_1,\infty)$}\\
    P(x) - \int_{y = \delta}^{x} \alpha \, dP(y) & \text{ for $x\in (x_0,\hat x)$}\\
   P(x) + \left( \frac{x - \hat{x}}{x_1 - \hat{x}} \right) \int_{y = \delta}^{\hat{x}} \alpha \, dP(y) & \text{ for  $x\in [\hat x,x_1)$}
    \end{cases}
\]
In particular, since $\alpha>0$,  we have $P(x)>Q(x)$ for $x\in(x_0,\hat x)$, and $P(x)<Q(x)$ for $x\in[\hat x,x_1)$ and $P(x)=Q(x)$ everywhere else. 
\end{proof}
    
\begin{proof}[\textup{\textbf{Proof of Proposition \ref{prop:cost characterization}}}]
It is clear that if $C_{F}$ can be expressed as in Equation (\ref{eq:cost_P}),
then it satisfies all conditions in Axiom \ref{axm:cost_P}. Therefore,
we proceed with the sufficiency of these conditions. Since $C_{F}(.)$
satisfies \textit{Linearity} ($iii$) and \textit{Continuity} ($v$) over $\bar {\mathcal E}$, by Herstein
and Milnor (1953), there exists some measurable continuous real valued
function $c_{F}$ such that $C_{F}(P)=\int_{\mathbb{R}}c_{F}(x)dP(x)$
for any $P\in \bar{\mathcal E}$. By \textit{prior-independence} ($ii$), the function
$c_{F}$ is independent of the distribution $F$, and so we write
$c$ instead of $c_{F}$. By \textit{Consistency} ($i$), we have $C_F(P)=C_F(P^{F})$
for any $P\in{\mathcal E}$. Thus, $C_F(P)=\int_{\mathbb{R}}c(x)dP^{F}(x)$.
Using the definition of $P^{F}$, we obtain $C_F(P)=\int_{\mathbb{R}}\left(\int_{\mathbb{R}}c(x)dP_{\theta}(x)\right)dF(\theta)$.

It remains to show that $c$ is strictly unimodal. Since for any $P$ and $\theta$, the  distribution $P_\theta$
is symmetric, we can assume without loss of generality that $c$ is
an even function, and so $c(-x)=c(x)$ for all $x\in\mathbb{R}$. We thus focus on the subdomain $\mathbb R_+$. Suppose for contradiction that there is an interval $[x_0,x_1]$
with $x_0<x_1$ such that $c$ is weakly increasing over
it with $c(x_0)<c(x_1)$. 
Fix $\hat x$ and let $P,Q\in \bar{\mathcal E}$ with $P\triangleright Q$ such that $P(x)=Q(x)$ for $x\in[0,x_0]\cup[x_1,\infty)$, $P(x)>Q(x)$ for $x\in(x_0,\hat x)$, and $P(x)<Q(x)$ for $x\in[\hat x,x_1)$. Such an experiment exists by Lemma \ref{lm:garbled Q exists}. Moreover, since $P(x_1)=Q(x_1)$, we have: 
\[
\int_{[x_0,\hat x]}[dP(x)-dQ(x)]=\int_{(\hat x,x_1]}[dQ(x)-dP(x)].
\]
As $c$ is increasing on $[x_0,x_1]$ by construction, we have $c(x)\leq c(y)$ for all $x\in[x_0,\hat x]$
and $y\in(\hat x,x_1]$, and therefore
\[
\int_{[x_0,\hat x]}c(x)[dP(x)-dQ(x)]\leq\int_{(\hat x,x_1]}c(x)[dQ(x)-dP(x)].
\]
It thus follows that
\begin{align*}
C_{F}(P)/2 & =\int_{[0, x_0)\cup(x_1,\infty)}c(x)dP(x)+\int_{[x_0,\hat x]}c(x)dP(x)+\int_{(\hat x,x_1]}c(x)dP(x)\\
 & \leq\int_{[0,x_0)\cup(x_1,\infty)}c(x)dQ(x)+\int_{[x_0,\hat x]}c(x)dQ(x)+\int_{(\hat x,x_1]}c(x)dQ(x)\\
 & =C_{F}(Q)/2,
\end{align*}
showing that $C_{F}(Q)\geq C_{F}(P)$, a contradiction.
Moreover, it is clear from previous arguments that $c$ cannot be
constant over any interval $[x_0,x_1]$. Thus, $c$ must
be strictly decreasing over $\mathbb{R}_{+}$, and so it must be strictly
increasing over $\mathbb{R}_{-}$ implying that $c$ is strictly unimodal at zero. 
\end{proof}
\end{toappendix}
Our characterization result demonstrates that, under the stated axioms, the cost of experiments takes the form of an expected cost, where the expectation is over a prior-independent noise cost function $c(x)$. The integral representation follows directly from the linearity and continuity of the cost function, mirroring the expected utility characterization in \cite{herstein1953axiomatic}. The uni-modality of the noise cost function $c(x)$ is a consequence of Blackwell-monotonicity. If $c(x)$ were not unimodal, we could construct an experiment $Q$ from another experiment $P$ by shifting probability mass from a region of higher cost to a region closer to the center with lower cost. This adjustment would make $Q$ more informative than $P$ while incurring a lower total cost, contradicting Blackwell-monotonicity.



\subsection{Benefits of information}
A \textit{decision rule} $\psi: S \to \mathcal{A}$ for the decision maker is a function that maps an observed signal $s\in S$ to an action $\psi(s) \in \mathcal{A}$. A decision $\psi(s)$ at signal $s$ is interim-optimal if it maximizes the decision-maker's expected payoff, conditional on the observed signal $s$. We want to characterize optimal decision rules in terms of the decision-maker's conditional expectation after observing a signal about the state. To obtained a unified representation of the posterior, we first apply the Lebesgue decomposition theorem to decompose $F_{P,s}$ into an absolutely continuous part (with respect to the Lebesgue measure) and a discrete part (captured by Dirac measures at jump points), and then represent 
$F_{P,s}$ using its Radon-Nikodym derivative relative to an appropriately chosen reference measure. Let $D=\{\theta|F(\theta_+)-F(\theta_-)>0\}$ be the set of discontinuities in $F$. Define the reference measure for $F$ as 
\[
\mu = \lambda + \sum_{\theta \in D}  d_{\theta},
\]
where $\lambda$ is the standard Lebesgue measure on $\mathbb R$ that tracks the absolutely continuous part of $F$ and $d_{\theta}$ denotes the Dirac measure at  $\theta$ that tracks the discrete jumps of $F$.  The Radon-Nikodym derivative $f(\theta)$ of $F$ with respect to $\mu$ is then given by
\[
f(\theta) = \frac{dF}{d\mu} =
\begin{cases}
f_c(\theta), & \text{ if }\theta\in \mathbb R\setminus D, \\
\Delta F(\theta^i), & \text{ if } \theta \in  D.
\end{cases}
\]
At points where $F$ is absolutely continuous, $f(\theta)$ corresponds to the density of the absolutely continuous part with respect to the Lebesgue measure. At points where $F$ has jumps, $f(\theta)$ gives the weight of the Dirac measure, i.e., the size of the jump $\Delta F(\theta^i)=F(\theta_+)-F(\theta)$.   Now,  consider the probability distribution  $\pi_{F,P}$ over the signal space that is induced by $F$ and $P$:
\begin{align}\label{eq:dF_H,s}
\pi_{F,P}(S)=\int_{\mathbb R}dP_{\theta}(\sigma_{\theta}^{-1}(S))f(\theta)d\mu(\theta), \quad\forall S\in \mathcal B
\end{align}
where $\sigma_\theta^{-1}(S)=\{x\in \mathbb R| \sigma(\theta,x)\in S\}$  denotes the pre-image of $S$ under $\sigma(\theta,\cdot)$. At any signal $s\in \supp \pi_{F,P}$, the measure element of the posterior distribution $dF_{P,s}$ is given by:
\[
dF_{P,s}(\theta) = \frac{dP_\theta(\sigma_\theta^{-1}(\{s\})) f(\theta) \, d\mu(\theta)}{\int_{\mathbb R} dP_{\hat{\theta}}(\sigma_{\hat{\theta}}^{-1}(\{s\})) f(\hat{\theta}) \, d\mu(\hat{\theta})}.
\] 
An interim-optimal decision-rule assigns to each signal $s\in \supp \pi_{F,P}$ an action $\psi^*_{F,P}(s)\in \mathcal A$, that maximizes her expected utility with respect to her posterior belief $F_{P,s}$ associated with $s$:
\begin{equation}
\psi^{*}_{F,P}(s)\in\arg\max_{a\in \mathcal{A}}\int_{\mathbb{R}}u(\theta,a)dF_{P,s}(\theta). \label{eq:bdr}
\end{equation}
Note that since $u(\cdot,\cdot)$ is continuous in both arguments, and $\mathcal{A}$ is compact, an optimal decision rule $\psi^{*}_{F,P}(s)$ exists for all $s\in \supp 
\pi_{F,P}$. We record this result for future reference. 
\begin{lemma}
    An optimal decision rule $\psi^{*}_{F,P}(s)$ exists for all $s\in \supp \pi_{F,P}$. 
\end{lemma}
We define the \textit{gross-benefit} of the decision-maker as the expected utility under the optimal decision-rule:
\begin{equation}
B_F(P)=\int_{s\in S}\left(\int_{\theta \in \mathbb{R}} u(\theta,\psi^{*}_{F,P}(s))dF_{P,s}(\theta)\right)\pi_{F,P}(s)ds.\label{eq:gbf}
\end{equation}
The following lemma confirms that the benefit function is continuous in the experiment. 
\begin{lemma}\label{lm:B continuous}
    The decision-maker's benefit function $B_F(P)$ is continuous in $P$. 
\end{lemma}
\begin{proof}
Notice that both $\pi_{F,P}(s)$ and $F_{P,s}$ are continuous functions
of $P$. Moreover, by the Maximum Theorem, for each $s\in S$, the indirect utility function $u(\theta,\psi^{*}_{F,P}(s))$
is also a continuous function of $P$. Therefore, the gross-benefit function $B_F(P)$ is a continuous function of $P$. 
\end{proof}

\section{Optimal Experiments}

We now turn to the decision-makers problem of selecting an optimal experiment. The decision-maker seeks to maximize her net-benefit function over a set of feasible experiments. Our analysis proceeds in three steps. First, we establish a key result: any experiment can be replaced by a uniform experiment without reducing net benefit. This insight, which we term “uniform dominance,” allows us to restrict the search for an optimal experiment to the set of uniform experiments. We then use this result to prove the existence of an optimal experiment. Third, we characterize the optimal uniform experiment and derive the necessary conditions for its construction.

The decision-maker’s objective is to find an experiment $P^* \in \mathcal{E}$ that maximizes her \textit{net-benefit}  function:
\begin{equation}
\max V_F(P)\equiv B_F(P)-C_F(P) \label{eq:VF}
\end{equation}
Note that since both $B_F$ and $C_F$ are continuous, the net-benefit function is continuous. The continuity of the decision-maker's net-benefit function $V_F$ is a central for the choice of optimal information structures, which is discussed in the next section.

The optimal experiment must trade-off the benefit and cost of noise.  On the one hand, a reduction in the noise of the signal shifts probability mass towards zero, which makes the signal more informative, but also leads to an increase in the cost of the experiment. On the other hand, increasing the noise in the signal spread out the probability mass and thus reduces overall cost, but making the signal less informative.


\subsection{Existence of an Optimal Experiment}
\nosectionappendix

The following example shows that we need restrictions on the space of  experiments to ensure the existence of an optimal experiment. Specifically, we show that we can construct a sequence of uniform experiments, such that the cost of the experiment vanishes in the limit, but at each step, the experiment perfectly reveals the state. So while it is possible to make the cost of perfectly learning the state arbitrarily small, reducing the cost to zero will force the information structure to be completely uninformative. 

For any $\delta\in[0,\infty)$ in the extended real-line, let $H_{\delta}$ denote the uniform distribution over $[-\delta,\delta]$. We call $\delta$ the \textit{noise-level}.  If an experiment $P$ is such that $P_\theta = H_{\delta(\theta)}$ for some $\delta(\theta)\in[0,\infty]$ for each $\theta$, we call it a \textit{uniform experiment}.  Denote the set of uniform experiments by $\mathcal{E}_u$. 

\begin{example}
 Consider a prior $F$ which assigns positive probability weight exclusively to two states $\theta_0$ and $\theta_1>\theta_0$, and let $p$ be the probability of $\theta_1$. For each $i=1,2,\ldots$, let $H^i=\{H_{\delta^i(\theta)}\}$ be a uniform experiment with $\delta^i(\theta_0)=\infty$ and $\delta^i(\theta_1)=d^i$, where where $d^1,d^2,\ldots$ is an increasing sequence with $d^i\to \infty$. Denote by $h_{\delta}(x)$ the density at any point $x\in \mathbb R$. Note that for the improper distribution $H_{\infty}$, we have 
 \[
 h_{\infty}=\lim_{d\to \infty}=\frac{1}{2d}=0.
 \]
 At step $i=1,2,\ldots$, the decision-maker's posterior belief regarding $\theta$ after observing a signal $s\in [q-d^i,q+d^i]$ is 
 \[
\Pr(\theta=\theta_1\mid s)=\frac{ph_{d^i}}{ph_{d^i}+(1-p)h_{\infty}}=1.
 \]
So this experiment perfectly reveals the state to the decision-maker at each step $i=1,2,\ldots$. On the other hand, the cost of this information structure tends to zero, since
 \[
 \lim_{i\to \infty}C_F(H^i)=\lim_{i\to \infty}\Bigl(p \lim_{d\to \infty}\int_{-d}^d c(x)dx+(1-p)\frac{1}{d^i}\int_{-d^i}^{d^i}c(x)dx\Bigr)=0
 \]
 Note, however, that the experiment that assigns infinite noise to both states $\theta_0$ and $\theta_1$ is entirely uninformative, which shows us that there is no optimal information structure in this case. 
\end{example}

To rule out the the existence problems that arise from allowing unbounded noise in experiments, we impose the restriction that the noise in experiments shall be bounded. Specifically, for any level $b>0$, denote by $\mathcal E^b$ the set of  all experiments $P=\{P_\theta\}$, where
\[
\sup \Bigl\{\bigcup_\theta \supp P_\theta\Bigr\}< b.
\]
In the space of bounded uniform experiments, we  show that an optimal experiment exists.\footnote{In general, to guarantee the existence of a maximizer, the domain of the choice problem given in equation \eqref{eq:VF} can be specified to be a compact set of admissible experiments.} 


\begin{lemma}[Existence of an Optimal Experiment]
\label{lm:exist}
For all $b\ge 0$, there exists an optimal  experiment $P^* \in \mathcal E^b$ such that  
\begin{equation}\label{eq:gbf2}
  V_F(P^*) \;=\; \max_{P \in \mathcal E^b} \Bigl( B_F(P) \;-\; C_F(P) \Bigr).  
\end{equation}
\end{lemma}

\begin{proof}
See Appendix.
\end{proof}

\begin{toappendix}
    \begin{proof}[\textup{\textbf{Proof of Lemma \ref{lm:exist}}}]
Fix $b>0$. Note that each $P\in\mathcal E^b$ is by construction bounded above by a common limit $b>0$. As such, the limit of any sequence of bounded experiments in $E^b$ will also be bounded by $b$. Hence, $\mathcal E^b$ is compact. Note further, that the decision-maker's net-benefit function, given by
$$
V_F(P) = B_F(P) - C_F(P), 
$$  
is continuous, since $B_F$ is continuous by Lemma \ref{lm:B continuous}, and the cost function $C_F$ is continuous by construction. Therefore, $V_F$ is continuous on $\mathcal{E}^b$. Then, by the extreme value theorem, a maximum of $V_F(P)$ over $\mathcal{E}^b$ must exist. 
\end{proof}
\end{toappendix}


\subsection{Uniform Dominance Principle}
\nosectionappendix

Next, we demonstrate that focusing solely on uniform experiments is without loss of generality. The following theorem shows that, given any (possibly non-uniform) experiment in $\mathcal{E}$, we can construct a uniform experiment in $\mathcal{E}_u$ that achieves at least the same net benefit. This is the key insight behind the following result:

\begin{theorem}[Uniform Dominance Principle]\label{thm:uniform}
For any experiment $P \in \mathcal{E}$, there exists a uniform experiment $H \in \mathcal{E}_u$ such that 
\[
V_F(H) \;\ge\; V_F(P).
\]
\end{theorem}

\begin{proof}
See Appendix.
\end{proof}

\begin{toappendix}
    \begin{proof}[\textup{\textbf{Proof of Theorem \ref{thm:uniform}}}]
The proof is given in several steps. Recall
that, by Assumption \ref{ass:regular_P}, for any $\theta\in\mathbb{R}$, the density $P_\theta'\equiv d P_{\theta}/dx$
is an even function which is strictly unimodal at $0$.  The uniform density on $[-\delta,\delta]$ is:
\[
H_\delta'(x) = 
\begin{cases} 
\frac{1}{2\delta}, & |x| \leq \delta, \\
0, & |x| > \delta.
\end{cases}
\]
We now construct a sequence of approximations to $P'_{\theta}$ for
each $\theta\in\mathbb{R}$ consisting of a weighted sum of uniform
distributions which lies below $P_{\theta}'$ on the real line, but
coincides with $P_{\theta}'$ along points in a grid.

\begin{enumerate}[($i$)]
\item Define the grid: For any $n=0,1,2,...$, let $k_{n}=3^{n}$ denote
the number of grid points, and let $\delta_{n}=\frac{1}{2^{n}}$ be
the grid spacing. Note that for $n\to\infty$, we have $\delta_{n}\to0$
and $\delta_{n}{\cdot}k_{n}=\frac{3^{n}}{2^{n}}\to\infty$, so in
the limit, the grid becomes infinitely fine and spans the entire real
line. 
\item Construct an approximation of $P_{\theta}'$ through uniform distributions:
Let $Q_{\theta,n}'$ be the density defined as the weighted sum of
uniform distributions over this grid. That is, let 
\[
Q_{\theta,n}'(x)=\sum_{j=1}^{k_{n}}\alpha_{n,j}^{\theta}H_{j\delta_{n}}'(x),
\]
where $\alpha_{n,j}^{\theta}\in(0,1)$ for $j=1,\ldots,k_{n}$. We
want to choose weights such that $Q_{\theta,n}'(x)=P_{\theta}'(x)$
at each $x=\pm j\delta$ for $j=1,2,\ldots,k$. Since $H_{i\delta_{n}}'(x)$
is nonzero only for $|x|\leq i\delta_{n}$, we note that $H_{i\delta_{n}}(j\delta_{n})$
contributes only when $j\leq i$. Specifically: 
\[
H_{i\delta_{n}}'(j\delta_{n})=\begin{cases}
\frac{1}{2i\delta_{n}}, & j\leq i,\\
0, & j>i.
\end{cases}
\]
Thus, at $x=j\delta_{n}$, the equation for $Q_{\theta,n}'(j\delta_{n})$
becomes: 
\begin{align}
P'(j\delta_{n})=\sum_{i=j}^{k_{n}}\frac{\alpha_{n,i}^{\theta}}{2i\delta_{n}}\label{eq:approx at grid points}
\end{align}
\item We show that such an approximation exists: Condition \eqref{eq:approx at grid points}
provides one equation for each $j=1,2,\ldots,k$. We write these equations
in matrix form. Define the upper triangular $k_{n}\times k_{n}$ matrix
$A_{n}$ with entries: 
\[
A_{n}^{ji}=\begin{cases}
\frac{1}{2i\delta}, & i\geq j,\\
0, & i<j.
\end{cases}
\]
Let 
\[
\boldsymbol{\alpha}_{n}^{\theta}=[\alpha_{n,1}^{\theta},\alpha_{n,2}^{\theta},\ldots,\alpha_{n,k_{n}}^{\theta}]^{T}
\]
denote the vector of weights, and let 
\[
\mathbf{b}_{n}=[P'(\delta),P'(2\delta),\ldots,P'(k_{n}\delta)]^{T}
\]
be the vector of values of $P'(x)$ at the specified boundary points.
Then, the system of equations can be written as: 
\[
A_{n}\boldsymbol{\alpha}_{n}^{\theta}=\mathbf{b}_{n}.
\]
Since $A_{n}$ is an upper triangular matrix, it has full rank and
thus is invertible, so that the linear system has a unique solution
given by 
\[
\boldsymbol{\alpha}_{n}^{\theta}=(A_{n})^{-1}\mathbf{b}_{n}.
\]
Therefore, the vector of weights $\boldsymbol{\alpha}_{n}^{\theta}$
exists and is uniquely determined. 
\item We show that the sum of weights is less than 1. Note that for all
$n$: $Q_{\theta,n}'(x)\le P_{\theta}'(x)$ and thus 
\[
\int_{\mathbb{R}}Q_{\theta,n}'(x)dx\le\int_{\mathbb{R}}P_{\theta}'(x)dx=1.
\]
Moreover: 
\[
\int_{\mathbb{R}}Q_{\theta,n}'(x)dx=\int_{\mathbb{R}}\left(\sum_{j=1}^{k_{n}}\alpha_{n,j}^{\theta}H_{j\delta_{n}}'(x)\right)dx=\sum_{j=1}^{k_{n}}\alpha_{n,j}^{\theta}.
\]
Together, it follows that 
\[
\sum_{j=1}^{k_{n}}\alpha_{n,j}^{\theta}\leq1.
\]
\item We now show that for each $n$, there is a uniform distribution which is
at least good as $Q_{\theta,n}$: Suppose $\psi_{P}^{*}$ is the optimal
decision rule under the experiment $P$. For each $n=0,1,...$ we
have 
\begin{align*}
W(Q_{\theta,n},\psi_{P}^{*}) & =\int_{\mathbb{R}}\left[u(\theta,\psi_{P}^{*}(\sigma_{\theta}(x))-c(x)\right]dQ_{n,\theta}(x)\\
 & =\sum_{j=1}^{k_{n}}\alpha_{n,j}^{\theta}\left(\int_{\mathbb{R}}\left[u(\theta,\psi_{P}^{*}(\sigma_{\theta}(x))-c(x)\right]dH_{j\delta_{n}}(x)\right)\\
 & \le\int_{\mathbb{R}}\left[u(\theta,\psi_{P}^{*}(\sigma_{\theta}(x))-c(x)\right]d{H}_{j_{n}(\theta)\delta_{n}}(x)\\
 & \leq\int_{\mathbb{R}}\left[u(\theta,\psi_{P}^{*}(\sigma_{\theta}(x))-c(x)\right]d{H}_{\delta(\theta)}(x)
\end{align*}
where we define 
\[
j_{n}(\theta)=\arg\max_{j\in\{0,\ldots,k_{n}\}\cup\{\infty\}}\int_{\mathbb{R}}\left[u(\theta,\psi_{P}^{*}(\sigma_{\theta}(x))-c(x)\right]dH_{j\delta_{n}}(x)
\]
for each $n=0,1,...$ and $\delta(\theta)=j_{n^{*}}(\theta)\delta_{n^{*}}$
with 
\[
n^{*}=\arg\max_{n\in\{0,1,,2\ldots\}\cup\{\infty\}}\int_{\mathbb{R}}\left[u(\theta,\psi_{P}^{*}(\sigma_{\theta}(x))-c(x)\right]dH_{j_{n}(\theta)\delta_{n}}(x)
\]
Now, define the uniform experiment $H=\{H_{\delta(\theta)}\}_{\theta}$.
By definition, we have 
\begin{align*}
V_{F}(H,\psi_{P}^{*}) & =\int_{\mathbb{R}}\left(\int_{\mathbb{R}}\left[u(\theta,\psi_{P}^{*}(\sigma_{\theta}(x))-c(x)\right]d{H}_{\delta(\theta)}(x)\right)dF(\theta)\\
 & \geq\int_{\mathbb{R}}W(Q_{\theta,n},\psi_{P}^{*})dF(\theta)=V_{F}(Q_{n},\psi_{P}^{*}).
\end{align*}
\item Take the limit: By construction, we have $Q_{n}\to P$, and so by
continuity of $V_{F}(.,\psi_{P}^{*})$, we have $V_{F}(Q_{n},\psi_{P}^{*})\to V_{F}(P,\psi_{P}^{*})=V_{F}(P)$
as $n\to\infty$. Moreover, the sequence $\{V_{F}(Q_{n},\psi_{P}^{*})\}$
is bounded above by the value $V_{F}(H,\psi_{P}^{*})$, and so $V_{F}(H,\psi_{P}^{*})\geq V_{F}(P)$.
Since by definition $V_{F}(H)\geq V_{F}(H,\psi_{P}^{*})$, we conclude
that $V_{F}(H)\geq V_{F}(P)$.
\end{enumerate}

\end{proof}
\end{toappendix}

The Uniform Dominance Principle states that for any statistical experiment, there exists a corresponding uniform experiment—a noise distribution that is uniform across some interval around zero—that provides the decision-maker with at least as high a net benefit. In the proof, which is found in the appendix, we show that one can approximate any noise distribution by a (finite) mixture of uniform distributions with appropriately chosen weights, and by linearity and Blackwell monotonicity of information costs, selecting a single best uniform component within that mixture never reduces the decision-maker’s net benefit. Thus, even starting with an arbitrary $P \in \mathcal{E}$, we do at least as well by moving to a uniform distribution in $\mathcal{E}_u$.

The Uniform Dominance Principle has a number of important implications. First, uniform dominance significantly simplifies the problem of optimal learning by reducing an infinite-dimensional optimization over complex noise distributions to a relatively simple one-dimensional problem. Instead of grappling with the full space of possible noise distributions, it suffices to find the  "spread" of optimal uniform noise for each state, which amount to identifying a single parameter. This reduction not only makes the problem mathematically tractable but also allows for clearer intuition about the trade-offs faced by decision-makers.

The Uniform Dominance Principle also provides a foundation for signals with uniform noise in theoretical models of information acquisition. A leading example in this context is the global games literature, following the work of \cite{carlsson1993global}. In this literature, players observe noisy signals about an underlying state where the noise is drawn uniformly from some bounded interval. Our findings suggest that uniform noise effectively captures the essential trade-off inherent in costly information acquisition. Importantly, in constrast to posterior-based cost models, which suffer from the free-at-full-information (FFI) problem, our approach directly applies to strategic settings. Consequently, whether in strategic environments like global games or other related contexts, uniform noise should not be viewed solely as a simplifying assumption. Instead, it may emerge naturally as the optimal choice of a decision-maker with only minimal constraints are placed on the cost structure of signal generation.

The Uniform Dominance Principle also provides a compelling justification for the use of uniform noise in econometric analysis, particularly in settings where noise is viewed as a decision variable or the result of optimization. In many applied contexts, data generation involves processes where noise reflects underlying trade-offs. For example, in experiments or surveys, researchers or participants may choose how much effort to exert in providing accurate responses, implicitly determining the level of noise in the data. Similarly, firms reporting financial data may strategically balance accuracy and cost, leading to structured patterns of noise. In these cases, the choice of noise is not arbitrary but reflects optimization under constraints such as cost, informativeness, or regulatory requirements. The Uniform Dominance Principle suggests that when noise is costly to manage, uniform noise may emerge as an optimal choice due to its balance between spreading information evenly and minimizing costs, reinforcing its relevance in these types of datasets.


\subsection{Characterization}

We now want to identify the characterizing conditions of a
solution for the decision-maker's optimization problem. Because of the Uniform Dominance Principle, there is no loss in focusing our attention uniform experiments. To this end, denote by $\mathcal E_u^b$, the space of bounded uniform experiments. For any $H\in \mathcal E_u^b$, let $\psi_{F,H}^{*}$ be a best decision
rule which solves the corresponding problem in (\ref{eq:bdr}) for each $s\in S$. As noted above, we must have 
\[
\psi_{F,H}^{*}(s)\in\arg\max_{a\in A}\int_{\mathbb{R}}u(\theta,a)dF_{H,s}(\theta),
\]
where $F_{H,s}$ denotes the posterior belief of the decision-maker
for each $s\in S$ derived in Equation \eqref{eq:dF_H,s}. 
When we substitute the best decision rule $\psi_{F,H}^{*}$ into the
choice problem (\ref{eq:gbf2}), we obtain the following:
\[
V_{F}^{*}=\max_{H\in{\mathcal E}_{u}^b}\int_{\mathbb{R}}\left(\int_{\mathbb{R}}[u(\theta,\psi_{F,H}^{*}(\sigma_{\theta}(x)))-c(x)]\,dH_{\theta}(x)\right)dF(\theta).
\]
For any uniform experiment, we can find $\delta(\theta)\in[0,\infty]$, such that $H_{\theta}=H_{\delta(\theta)}$. We can thus rewrite the above
problem as
\[
V_{F}^{*}=\max_{\{\delta(\theta)\le b\}_\theta}\int_{\mathbb{R}}\left(\int_{-\delta(\theta)}^{+\delta(\theta)}[u(\theta,\psi_{F,H}^{*}(\sigma_{\theta}(x)))-c(x)]\,\frac{1}{2\delta(\theta)}dx\right)dF(\theta).
\]
Note that the choice of decision rule and experiment is relevant only on the support of $F$, and can be specified arbitrarily on sets with zero probability.

In order to derive a set of necessary conditions for identifying an optimal experiment, we now consider the effects of marginal changes in the bounds of each uniform distribution on the decision-makers expected payoff. Notice that by the Envelope Theorem, the indirect marginal effects of a
change in $\delta(\theta)$ on any state $\theta'\neq \theta$ through the best decision rule $\psi_{F,H}^{*}$ will be
null. Thus, we only need to consider the direct effects of marginal changes in $\delta$ on the decision-maker's payoff. Let
\[
W_{F,H,\theta}(\delta)=\int_{-\delta}^{+\delta}[u(\theta,\psi_{F,H}^{*}(\sigma_\theta(x)))-c(x)]\,\frac{1}{2\delta}dx. 
\]
denote the marginal value-contribution to the decision-maker's payoff in state $\theta\in \mathbb R$ as a function of the noise-level of the uniform distribution centered at zero. The function $W_{F,H,\theta}(\delta)$ is continuous in $\delta$, and in fact is Lipschitz-continuous as the following Lemma verifies.


\begin{lemma}\label{lm:W is lipschitz} $W_{F,H,\theta}$ is locally
Lipschitz continuous on $(0,\infty)$.
    
\end{lemma}

\begin{proof}
See Appendix.
\end{proof}

\begin{toappendix}
    \begin{proof}[\textup{\textbf{Proof of Lemma \ref{lm:W is lipschitz}}}]
Let $v_{\theta}(x)=u(\theta,\psi_{F,H}^{*}(\sigma(\theta,x)))-c(x)$
in short. Let $v_{\theta}^{max}=\max_{x\in\mathbb{R}}v_{\theta}(x)$
and $v_{\theta}^{min}=\min_{x\in\mathbb{R}}v_{\theta}(x)$. Let $B\subset(0,\infty)$
such that $\inf B>0$ and let $\delta_{1},\delta_{2}\in B$ such that
$\delta_{1}>\delta_{2}$. We have
\begin{align*}
|W_{F,H,\theta}(\delta_{1})-W_{F,H,\theta}(\delta_{2})| & =|\frac{1}{2\delta_{1}}\int_{-\delta_{1}}^{+\delta_{1}}v_{\theta}(x)dx-\frac{1}{2\delta_{2}}\int_{-\delta_{2}}^{+\delta_{2}}v_{\theta}(x)dx|\\
 & \leq A^{+}(\delta_{1},\delta_{2})+A^{-}(\delta_{1},\delta_{2})
\end{align*}
where 
\[
A^{+}(\delta_{1},\delta_{2})=|\frac{1}{2\delta_{1}}\int_{0}^{+\delta_{1}}v_{\theta}(x)dx-\frac{1}{2\delta_{2}}\int_{0}^{+\delta_{2}}v_{\theta}(x)dx|
\]
 and 
\[
A^{-}(\delta_{1},\delta_{2})=|\frac{1}{2\delta_{1}}\int_{-\delta_{1}}^{0}v_{\theta}(x)dx-\frac{1}{2\delta_{2}}\int_{-\delta_{2}}^{0}v_{\theta}(x)dx|.
\]
We have 
\begin{align*}
A^{+}(\delta_{1},\delta_{2}) & =|\frac{1}{2\delta_{1}}\int_{+\delta_{2}}^{+\delta_{1}}v_{\theta}(x)dx-(\frac{1}{2\delta_{2}}-\frac{1}{2\delta_{1}})\int_{0}^{+\delta_{2}}v_{\theta}(x)dx|\\
 & \leq|\frac{\delta_{1}-\delta_{2}}{2\delta_{1}}v_{\theta}^{max}-\frac{\delta_{1}-\delta_{2}}{2\delta_{1}}v_{\theta}^{min}|=\frac{\delta_{1}-\delta_{2}}{2\delta_{1}}(v_{\theta}^{max}-v_{\theta}^{min}).
\end{align*}
A similar argument shows that $A^{-}(\delta_{1},\delta_{2})\leq\frac{\delta_{1}-\delta_{2}}{2\delta_{1}}(v_{\theta}^{max}-v_{\theta}^{min})$.
Therefore, we have 
\begin{align*}
|W_{F,H,\theta}(\delta_{1})-W_{F,H,\theta}(\delta_{2})| & \leq\frac{\delta_{1}-\delta_{2}}{\delta_{1}}(v_{\theta}^{max}-v_{\theta}^{min})\\
 & \leq(\delta_{1}-\delta_{2})\frac{v_{\theta}^{max}-v_{\theta}^{min}}{\inf B}\\
 & =(\delta_{1}-\delta_{2})K_{\theta,B}
\end{align*}
where $K_{\theta,B}=\frac{v_{\theta}^{max}-v_{\theta}^{min}}{\inf B}$
implying that $W_{F,H,\theta}$ is locally Lipschitz continuous on
$B$. Notice that we have $(0,\infty)=\cup_{x\in(0,1)}(x,1/x)$, and
so $W_{F,H,\theta}$ is locally Lipschitz continuous on $(0,\infty)$.
    \end{proof}
\end{toappendix}


By Rademacher's theorem, Lipschitz continuity implies that $W_{F,H,\theta}$ is differentiable almost everywhere. It may fail to be differentiable at isolated points at which the decision rule $\psi^*_{F,H}$, which underlies $W_{F,H,\theta}(\delta)$, has jumps. Therefore, for the purpose of characterizing the optimal uniform experiment, we have to account for kinks in the objective function and thus cannot rely on standard first-order condition. 

To address this, we use the \textit{Clarke generalized derivative}, which extends first-order conditions to non-smooth functions by considering the "slopes" from all directions and combining them into a set. In this framework, the first-order condition states that a point is a candidate for a local extremum if zero lies within the set of these generalized slopes. Formally, let $E_{F,H,\theta}$ denote the set of points in $(0,\infty)$ at which
$W_{F,H,\theta}(\cdot)$ fails to be differentiable. The Clarke generalized derivative at $\delta$ is given by
\[
\partial_{C}W_{F,H,\theta}(\delta)=\operatorname{co}
\left[\lim_{\delta\downarrow\delta^{*}(q)}W'_{F,H,\theta}(\delta),\lim_{\delta\uparrow\delta^{*}(q)}W'_{F,H,\theta}(\delta)\right].
\]
Intuitively, the Clarke generalized derivative at a given point $\delta$ comprises the convex hull of the limit of sequence of derivatives at $\delta$ when approaching this point from either direction. 

\begin{example}
To illustrate this, consider the specific case $W(\delta)=W_{F,H,\theta}(\delta)=1/(|\delta|+a)$ for any number $a\ge 0$. For $\delta > 0$, $W(\delta)$ is differentiable with $W'(\delta) = -1/(\delta + a)^2$. For $\delta < 0$, we have $W'(\delta) = 1/(\delta - a)^2$. For $\delta \neq 0$, the Clarke generalized derivative equals the standard derivative, with $\partial_C W(\delta) = \{-1/(\delta + a)^2\}$ if $\delta > 0$, and $\partial_C W(\delta) = \{\frac{1}{(-\delta + a)^2}\}$ if $\delta < 0$. At $\delta = 0$, $W(\delta)$ is non-differentiable. Approaching $\delta = 0$ from the right gives $\lim_{\delta \to 0^+} W'(\delta) = -1/a^2$, and from the left, $\lim_{\delta \to 0^-} W'(\delta) = 1/a^2$. Thus, the Clarke generalized derivative is the interval from $-1/a^2$ to $1/a^2$, i.e., $\partial_C W(0) = \left[ -1/a^2, 1/a^2 \right]$. Note in particular that $0\in \partial_C W(0) = \left[ -1/a^2, 1/a^2 \right]$, meaning that there is a horizontal line which goes through $W(0)$ and lies otherwise above the graph of $W(\cdot)$. The condition $0\in \partial_C W(0) $ thus generalizes the standard first order-condition, and here, it corresponds to the fact that the function $1/(|\delta|+a)$ has a global maximum at 0, but is not differentiable at that point.
\end{example}


Note that since $W_{F,H,\theta}$ is Lipschitz-continuous and thus differential almost everywhere, 
$\partial_C W_{F,H,\theta}(\delta)$ is nonempty, and its elements are bounded by the Lipschitz constant of $W_{F,H,\theta}$. 
 We can now state the following result:


\begin{theorem} \label{thm:char}
Suppose $H^*=\{H_{\delta(\theta)}\}$ is a uniform experiment in $\mathcal E_u^b$. Given the optimal decision rule $\psi_{F,H}^{*}$, for each $\theta \in supp(F)$ we either have $(i)$ $0\in\partial_C W_{F,H,\theta}(\delta(\theta))$ and $\delta(\theta)< b$  or $(ii)$ $\delta(\theta)=b$.
\end{theorem}

\begin{proof}
See Appendix.
\end{proof}

\begin{toappendix}


\begin{proof}[\textup{\textbf{Proof of Theorem \ref{thm:char}.}}] For ease of notation, we suppress $F$ and $H$ from the subscripts. From Exercise 10.7a in Clarke's book it follows that if a function
$f:X\to\mathbb{R}$ is locally Lipschitz continuous, then $0\in\partial_{C}f(x)$
whenever $f$ has a local extremum at $x$ (Fermat's rule for stationarity
points). We now show that this claim is true for $W_{\theta}$ whenever
it has a local maximum at some $\delta\in(0,\infty)$. It is enough to show that if $\delta(\theta)<b$, then $0\in \partial_CW_{\theta}(\delta(\theta))$. By Theorem
10.27 in Clarke, we have 
\[
\partial_{C}W_{\theta}(\delta)=co\{\lim_{i\to\infty}W'_{\theta}(\delta_{i}):\delta_{i}\to\delta,\delta_{i}\notin E_{W_{\theta}}\},
\]
where $E_{W_{\theta}}$ is the set of points in $(0,\infty)$ at which
$W_{\theta}$ fails to be differentiable. By Rademacher's theorem,
the set $E_{W_{\theta}}$ has measure zero; that is, since $W_{\theta}$
is locally Lipschitz continuous on $(0,\infty)$, it is differentiable
almost at every point in $(0,\infty)$. Let $\{\delta_{i}\}\subset(0,\delta]\setminus E_{W_{\theta}}$
be a sequence of points such that $\delta_{i}\to\delta$. There must
be a convergent subsequence of points $\{\delta_{i_{k}}\}\subset\{\delta_{i}\}$
such that $W'_{\theta}(\delta_{i_{k}})\geq0$. Otherwise, we can always
find a point $\delta'<\delta$ arbitrarily close to $\delta$ such
that $W_{\theta}(\delta')>W_{\theta}(\delta)$ contradicting that
$\delta$ is a local maximum. Since $W'_{\theta}$ is bounded (due
to the local Lipschitz continuity), without loss of generality we
have $a=\lim_{i_{k}\to\infty}W'_{\theta}(\delta_{i_{k}})\in\partial_{C}W_{\theta}(\delta)$
such that $a\geq0$. A similar argument shows that we can find a convergent
sequence of points $\{\delta_{i_{k}}\}\subset[\delta,\infty)\setminus E_{W_{\theta}}$
such that $b=\lim_{i_{k}\to\infty}W'_{\theta}(\delta_{i_{k}})\in\partial_{C}W_{\theta}(\delta)$
with $b\leq0$. Since $0\in[b,a]$ we conclude that $0\in\partial_{C}W_{\theta}(\delta)$.

Notice that using the above equivalence, we can define the generalized
gradient at $0$ as $\partial_{C}W_{\theta}(0)=co\{\lim_{i\to\infty}W'_{\theta}(\delta_{i}):\delta_{i}\to0^{+},\delta_{i}\notin E_{W_{\theta}}\}$. Clearly, if $0$ is a local maximum, then
$\partial_{C}W_{\theta}(0)\cap(0,\infty)=\emptyset$.
\end{proof}
\end{toappendix}

The theorem provides as necessary condition for optimality. It says that if $\delta(\theta)$ is the noise-level of the optimal uniform distribution in state $\theta$, then the generalized gradient at $\delta(\theta)$ contains zero.  A key economic insight from the characterization result is that the optimal noise width $\delta(\theta)$ in each state must balance the trade-off between cost and informativeness so finely that any small adjustment in $\delta(\theta)$ would not improve the decision-maker’s net payoff. Concretely, the requirement $0 \in \partial_C W_{F,H,\theta}(\delta(\theta))$ means that if the noise interval were slightly expanded or contracted, the incremental benefit from additional information would be exactly matched by the incremental cost. This pins down the marginal condition for optimal uniform noise: the uniform distribution around each state $\theta$ is chosen so that increasing precision (i.e., shrinking the interval) no longer yields a net benefit, but also so that further “de-precision” (expanding the interval) would be too costly in terms of lost informativeness.

\bibliography{references}
\bibliographystyle{chicago}
\end{document}